\documentclass[aps,preprint,prl,superscriptaddress,showpacs,amsmath,amssymb]{revtex4}

\usepackage{graphicx,color}
\usepackage{dcolumn}
\newcolumntype{.}{D{.}{.}{-1}}
\usepackage{bm}
\usepackage{epstopdf}
\usepackage{subfigure}
\usepackage{textcomp}

\newcommand{\lfp}{LiFePO$_4$}

\newcommand{\ts}{$T^*$}
\newcommand{\tss}{$T^*_{\rm m}$}

\newcommand{\mb}{\(\mu _{\rm B}\)}

\newcommand{\tn}{$T_{\rm N}$}
\newcommand{\etal}{\textit{et~al.}}

\newcommand{\cp}{$c_{\rm p}$}

\newcommand{\cpm}{$c_{\rm p}^{\rm mag}$}

\newcommand{\axy}{$\alpha_{xy}$}

\newcommand{\jmk}{J/(mol\,K)}

\newcommand{\ala}{$\alpha_{\rm a}$}
\newcommand{\alb}{$\alpha_{\rm b}$}
\newcommand{\alc}{$\alpha_{\rm c}$}
\newcommand{\alv}{$\alpha_{\rm v}$}

\newcommand{\alm}{$\alpha_{\rm p}^{\rm mag}$}

\newcommand{\bsf}{$B_{\rm SF}$}
\newcommand{\bs}{$B_{\rm sat}$}

\newcommand{\bco}{$B_{\rm C1}$}

\begin{document}

\title{High-magnetic field phase diagram and failure of magnetic Gr\"{u}neisen scaling in LiFePO$_4$ }

\author{J.~Werner  \footnotemark[2]}
\email[Email:]{johannes.werner@kip.uni-heidelberg.de}
\affiliation{Kirchhoff Institute of Physics, Heidelberg University, INF 227, D-69120 Heidelberg, Germany}
\author{S.~Sauerland \footnotemark[2] \footnotetext[2]{Both authors contributed equally.}}
\affiliation{Kirchhoff Institute of Physics, Heidelberg University, INF 227, D-69120 Heidelberg, Germany}
\author{C.~Koo}
\affiliation{Kirchhoff Institute of Physics, Heidelberg University, INF 227, D-69120 Heidelberg, Germany}
\author{C.~Neef}
\affiliation{Kirchhoff Institute of Physics, Heidelberg University, INF 227, D-69120 Heidelberg, Germany}
\author{A.~Pollithy}
\affiliation{Kirchhoff Institute of Physics, Heidelberg University, INF 227, D-69120 Heidelberg, Germany}
\author{Y.~Skourski}
\affiliation{Dresden High Magnetic Field Laboratory (HLD-EMFL), Helmholtz-Zentrum Dresden Rossendorf, D-01314 Dresden, Germany}
\author{R.~Klingeler}
\affiliation{Kirchhoff Institute of Physics, Heidelberg University, INF 227, D-69120 Heidelberg, Germany}
\affiliation{Centre for Advanced Materials (CAM), Heidelberg University, INF 225, D-69120 Heidelberg, Germany}

\footnotetext[2]{Both authors contributed equally.}


\date{\today}

\begin{abstract}
We report the magnetic phase diagram of single-crystalline \lfp\ in magnetic fields up to 58~T and present a detailed study of magneto-elastic coupling by means of high-resolution capacitance dilatometry. Large anomalies at \tn\ in the thermal expansion coefficient $\alpha$ imply pronounced magneto-elastic coupling. Quantitative analysis yields the magnetic Gr\"{u}neisen parameter $\gamma_{\rm mag}=6.7(5)\cdot 10^{-7}$~mol/J. The positive hydrostatic pressure dependence $dT_{\rm N}/dp = 1.46(11)$~K/GPa is dominated by uniaxial effects along the $a$-axis. Failure of Gr\"{u}neisen scaling below $\approx 40$~K, i.e., below the peak temperature in the magneto-electric coupling coefficient [\onlinecite{toft2015anomalous}], implies several competing degrees of freedom and indicates relevance of recently observed hybrid excitations~[\onlinecite{yiu2017hybrid}]. A broad and strongly magnetic-field-dependent anomaly in $\alpha$ in this temperature regime highlight the relevance of structure changes. Upon application of magnetic fields $B||b$-axis, a pronounced jump in the magnetisation implies spin-reorientation at $B_{\rm SF} = 32$~T as well as a precursing phase at 29~T and $T=1.5$~K. In a two-sublattice mean-field model, the saturation field $B_{\rm sat,b} = 64(2)$~T enables the determination of the effective antiferromagnetic exchange interaction $J_{\rm af} = 2.68(5)$~meV as well as the anisotropies $D_{\rm b} = -0.53(4)$~meV and $D_{\rm c} = 0.44(8)$~meV.
\end{abstract}

\maketitle

\section{Introduction}

In addition to exceptionally high applicability of lithium orthophosphates~\cite{padhi1997phospho,chung2011electronically,park2010review} for electrochemical energy storage in Li-ion batteries, competing magnetic interactions, magnetic anisotropy and coupling of spin and electric degrees of freedom yield complex magnetic behaviour in  Li$M$PO$_4$ ($M$ = Mn, Fe, Co, Ni). The rich resulting physics is, e.g., demonstrated by ferrotoroidicity in LiCoPO$_4$ and LiNiPO$_4$ \cite{van2007observation,van2008anisotropy,zimmermann2009anisotropy}. In general, depending on the actual transition metal, Li$M$PO$_4$ develops long-range antiferromagnetic order at low temperatures and exhibits a large magneto-electric effect in the magnetically ordered phase~\cite{vaknin2004commensurate,rivera1994linear,toft2015anomalous}. The known magnetic phase diagrams of this family are rather complex, featuring incommensurate spin configurations, frustration, and usual magnetic excitations~\cite{vaknin2004,jensen2009,baek2014,rudisch2013,fogh2017magnetic,toft2012magnetic,toft2011high}.

Magnetic phase diagrams have been reported for all lithium orthophosphates~\cite{fogh2017magnetic,toft2012magnetic,toft2011high} except for LiFePO$_4$. At $B=0$~T, \lfp\ develops long-range antiferromagnetic order of $S=2$ spins of the magnetic Fe$^{2+}$-ions below \tn\ = 50~K~\cite{santoro1967antiferromagnetism}. The ordered moment amounts to 4.09~\mb ~\cite{rousse2003magnetic,toft2015anomalous} and the spins are mainly directed along the crystallographic $b$-axis (space group $Pnma$)~\cite{rousse2003magnetic}. Notably, the ground state features a collinear rotation of the spins towards the $a$-axis as well as spin canting along the $c$-axis with an overall rotation of the ordered moments of 1.3(1)$^\circ$ off the $b$-axis~\cite{toft2015anomalous,yiu2017hybrid}. The observed spin canting suggests the presence of Dzyaloshinsky-Moriya (DM) interactions which may account for the magneto-electric coupling in \lfp. In particular, as spin canting is not compatible with $Pnma$ symmetry, a lower crystal symmetry might appear below \tn ~\cite{li2006spin,toft2015anomalous}. Even in the absence of spin canting, an alternative mechanism to the ME effect may originate from orbital magnetic moments responding to polar distortions induced by an applied electric field~\cite{scaramucci2012linear}. Magnetic interactions have been studied by various groups using inelastic neutron scattering (INS) which imply competing antiferromagnetic interactions of however contradicting magnitude~\cite{li2006spin,toft2015anomalous,yiu2017hybrid}. When the INS data are analyzed including single-ion anisotropy which is strongly suggested by the results presented at hand, the dominating magnetic exchange is found in the $bc$-direction, i.e., $J_\mathrm{bc}\approx 0.46$ and 0.77~meV, respectively, which is by a factor of 2 - 4 smaller than $D$.~\cite{toft2015anomalous,yiu2017hybrid} Notably, rather dispersionless low-energy excitations have been found to persist up to 720~K which are discussed in terms of single-ion spin splitting.~\cite{yiu2017hybrid}


Here we report the magnetic phase diagram and magneto-elastic coupling in \lfp . Pronounced anomalies in the thermal expansion coefficients as well as pulsed-field magnetisation data are used to construct the magnetic phase diagram. The data imply spin-reorientation at $B_{\rm SF}\|b = 32$~T as well as a precursing phase at 29~T. $B_{\rm SF}$ and \bs\ are discussed in a two-sublattice mean-field model which yields effective antiferromagnetic exchange interaction $J_{\rm af} = 2.68(5)$~meV and anisotropies $D_{\rm b} = -0.53(4)$~meV and $D_{\rm c} = 0.44(8)$~meV. High-resolution dilatometry enables detailed studies of the interplay of spin, structure, and dielectric degrees of freedom. The magnetic Gr\"{u}neisen parameter is determined as $\gamma_{\rm mag}=6.7(5)\cdot 10^{-7}$~mol/J for $T\geq 40$~K. At lower temperatures, failure of Gr\"{u}neisen scaling indicates relevance of electric and/or structure degrees of freedom. Notably, a broad feature in the thermal expansion coefficient in the same temperature range further demonstrates the intimate coupling of spin, charge, and structure in \lfp .

\section{Experimental}

Single crystals of \lfp\ were grown by the high-pressure optical floating-zone method as reported in detail in Ref.~\onlinecite{neef2017high}. Magnetisation in static magnetic fields up to 5~T was studied by means of a Quantum Design MPMS-XL5 SQUID magnetometer and in fields up to 15~T in a home-built vibrating sample magnetometer (VSM)~\cite{Klingeler2006Magnetism}. Pulsed-magnetic-field magnetisation was studied up to 58~T at Helmholtz Zentrum Dresden Rossendorf by the induction method using a coaxial pick-up coil system~\cite{skourski2011high}. The pulse raising time was 7~ms. The pulsed-field magnetisation data were calibrated using static magnetic field measurements. The relative length changes $dL/L$ were studied on a cuboidally-shaped crystal with a dimension of 3x3x2~mm$^3$. The measurements were done by means of a three-terminal high-resolution capacitance dilatometer.~\cite{werner2017anisotropy} In order to investigate the effect of magnetic fields, the linear thermal expansion coefficients $\alpha_i = 1/L_i\cdot dL_i(T)/dT$ were studied in magnetic fields up to 15~T which were applied along the direction of the measured length changes $i=a, b, c$. In addition, the field-induced length changes were measured at various fixed temperatures in magnetic fields up to 15~T and the longitudinal magnetostriction coefficient $\lambda_i = 1/L_i\cdot dL_i(B_i)/dB_i$ was derived.

\section{Experimental results}

\subsection{Static magnetic susceptibility}

\begin{figure}	
\includegraphics[width=1.0\columnwidth,clip] {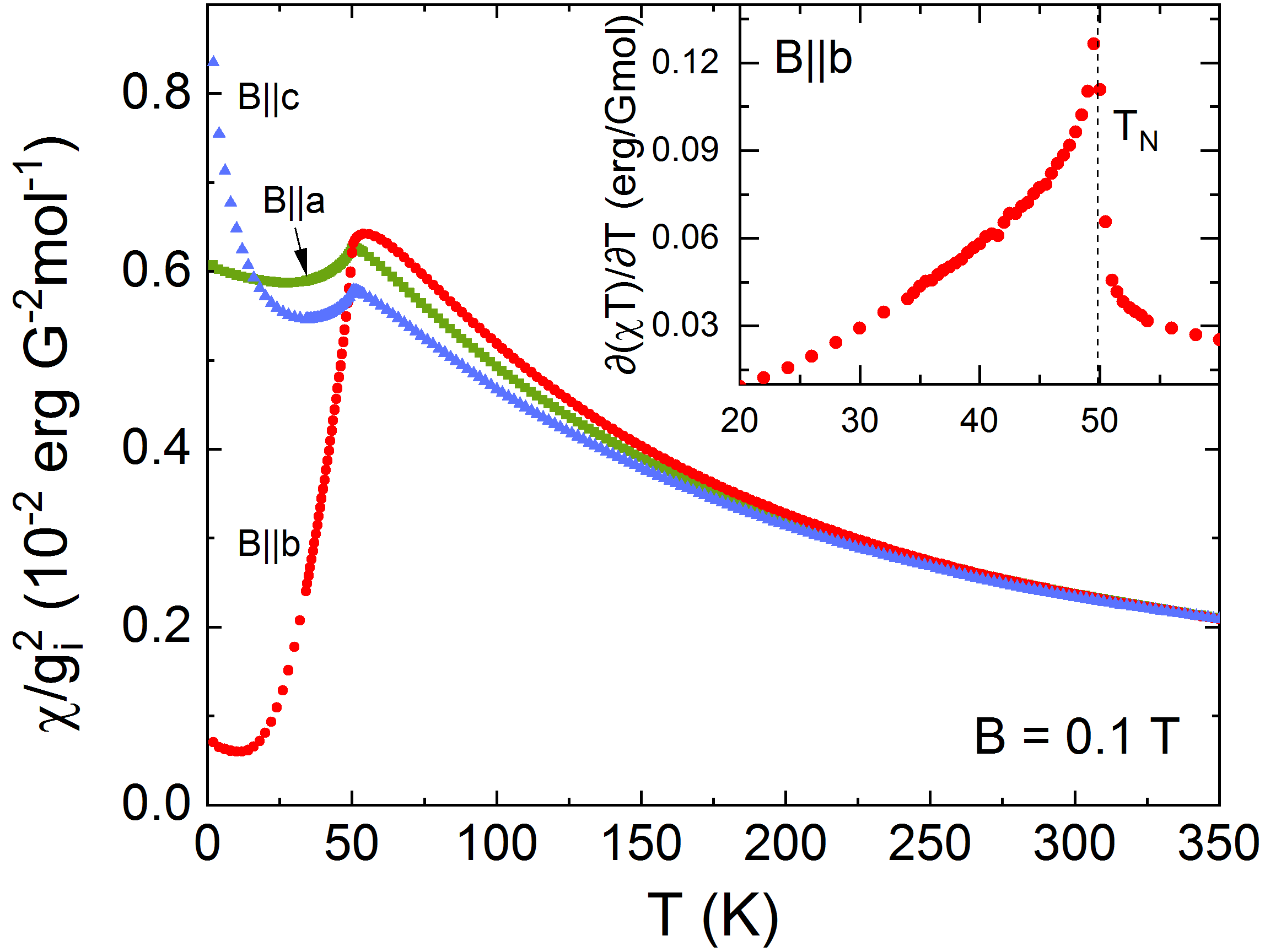}
\caption{Static magnetic susceptibility $\chi = M/B$ of \lfp\ vs. temperature for $B=0.1$~T applied along the three crystallographic axes. The data have been normalised by the respective components of the $g$-tensor as fitted to the high temperature behaviour (see the text). Inset: Corresponding derivatives $\partial (\chi T)/ \partial T$.}\label{chi}
\end{figure}

The onset of long-range antiferromagnetic order in \lfp\ at \tn = 50.0(5)~K is associated with pronounced anomalies in the magnetic susceptibility and in the thermal expansion (Figs.~\ref{chi} and \ref{alpha}). The magnetic susceptibility implies that the crystallographic $b$-axis is the easy magnetic axis in agreement with previous studies.~\cite{santoro1967antiferromagnetism,rousse2003magnetic} At high temperatures, the magnetic susceptibility obeys Curie-Weiss behaviour and the differences in magnetisation along the crystallographic axes can be associated to the anisotropy of the $g$-tensor. The data have hence been corrected by respective values of the $g$-factors, $g_{\rm i}$, which have been obtained by fitting the volume susceptibility by means of a Curie-Weiss-law and obtaining best overlap of $\chi_{\rm i}$ at high temperatures. This procedure yields $g_{\rm a} = 2.24(3)$, $g_{\rm b} = 2.31(2)$ and $g_{\rm c} = 1.99(3)$. However, the data imply anisotropy below 250~K which is not associated with the $g$-tensor, as visualized by Fig.~\ref{chi}.~\footnote{Correlation effects may extend to even higher temperatures which would affect the obtained $g$-values. Independent of that, magnetic anisotropy beyond the $g$-tensor extends to at least 250~K.} We also note a Curie-like upturn at low temperatures which is particularly pronounced for $B\|c$-axis, thereby indicating the presence of anisotropic quasi-free magnetic moments (cf. also Fig.~\ref{fig2}a).

\subsection{Thermal expansion}

\begin{figure}	
\includegraphics[width=1.0\columnwidth,clip] {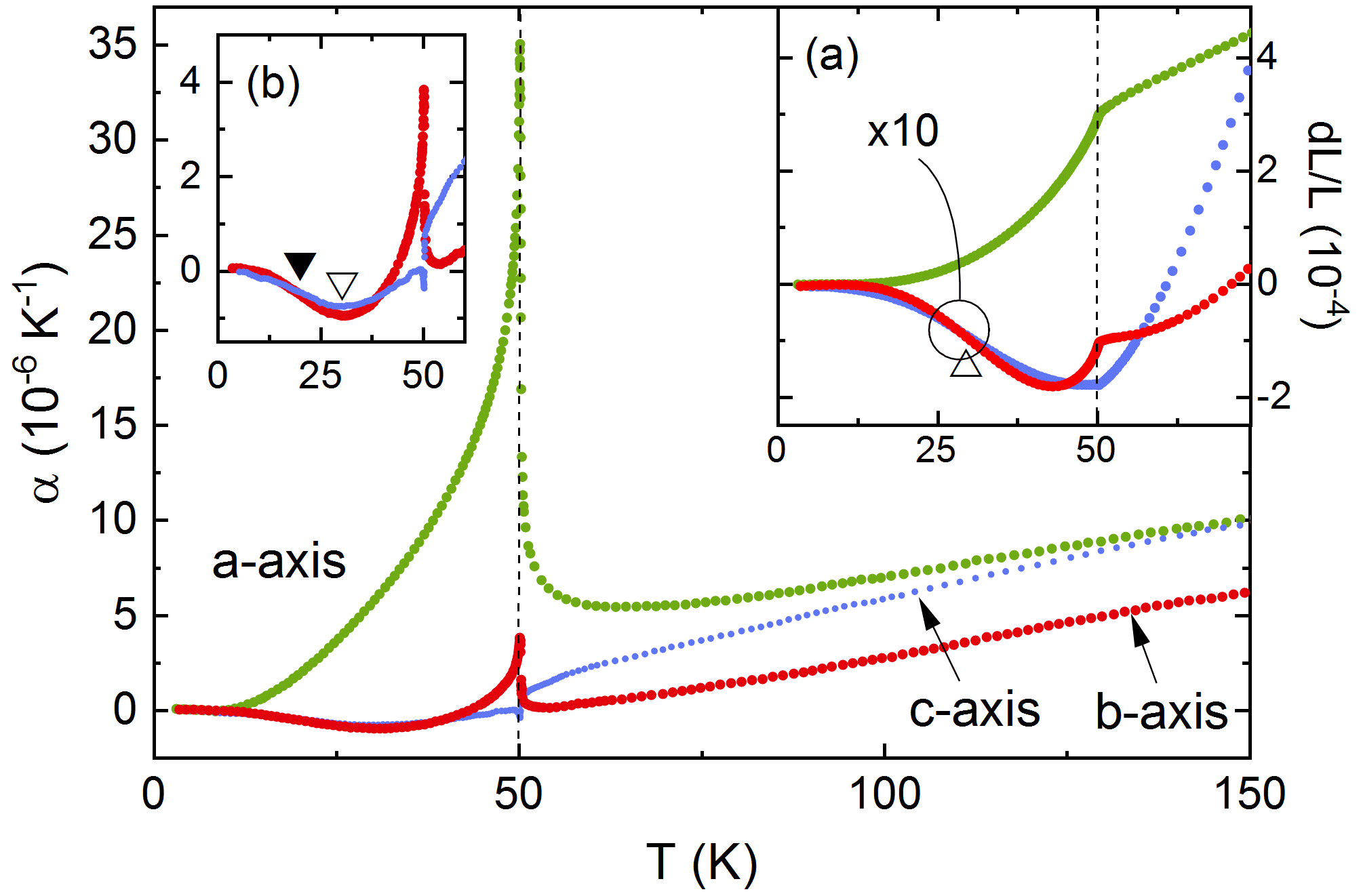}
\caption{Thermal expansion coefficient $\alpha$ along the three crystallographic axes. The dashed line shows \tn . Insets: (a) Associated relative length changes $dL/L$. Data for the $b$- and $c$-axis have been multiplied by 10. (b) Thermal expansion coefficient along the $b$- and $c$-axis up to 60~K. Open (closed) triangles label the temperatures \tss\ (\ts ) of a minimum (step) in $\alpha$ (see text).}\label{alpha}
\end{figure}

\begin{figure*}
\includegraphics[width=\textwidth]{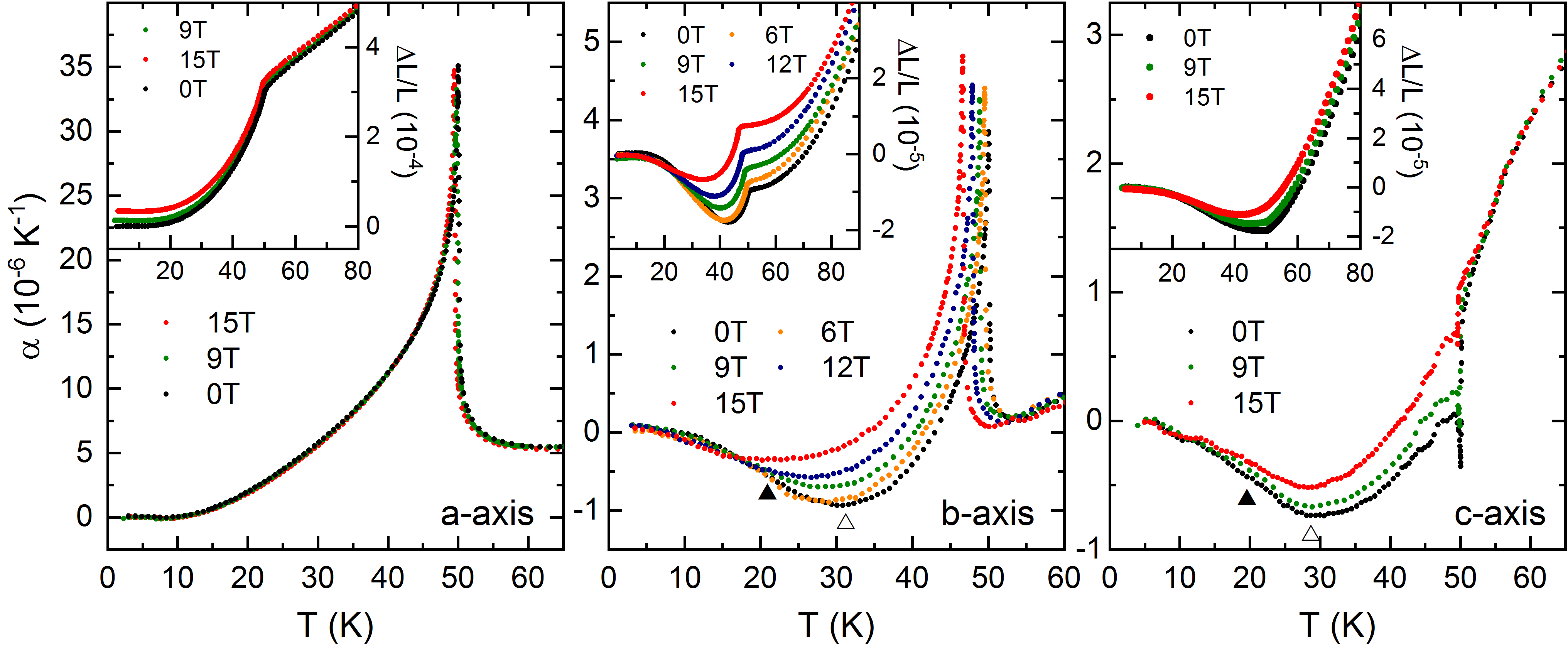}
\caption{Thermal expansion coefficient $\alpha$ at magnetic fields between 0 T and 15 T for all three crystallographic directions of \lfp. Insets show the corresponding length changes. Open (closed) triangles label the temperatures \tss\ (\ts ) of a minimum (step) in $\alpha$ (see text).} \label{fig:LT}
\end{figure*}

The evolution of long-range magnetic order at \tn\ is confirmed by sharp anomalies of the uniaxial thermal expansion coefficients $\alpha_{\rm i}$ ($i=a,b,c$) (Fig.~\ref{alpha}). The $\lambda$-shaped anomalies confirm the continuous nature of the phase transition. The measured length changes $dL/L$ shown in Fig.~\ref{alpha}a signal shrinking of the $a$- and $b$-axis upon evolution of long-range magnetic order at \tn\, while there is a slight increase of the $c$-axis. The anomalies confirm significant magneto-elastic coupling in \lfp . The signs of the anomalies show positive uniaxial pressure dependence of \tn\ for pressure along the $a$- and $b$-axis, i.e., $\partial T_{\rm N}/\operatorname{\partial p_i}>0$ for $i=a,b$. On the other hand, there is only a tiny anomaly in \alc\ indicating $\partial T_{\rm N}/\operatorname{\partial p_c}$ being negative and small. \tn\ also shows significant positive hydrostatic pressure dependence as shown by a very large anomaly of the volume thermal expansion coefficient (see the inset of Fig.~\ref{fig:grueneisen}).

\begin{figure}	
\includegraphics[width=1.0\columnwidth,clip] {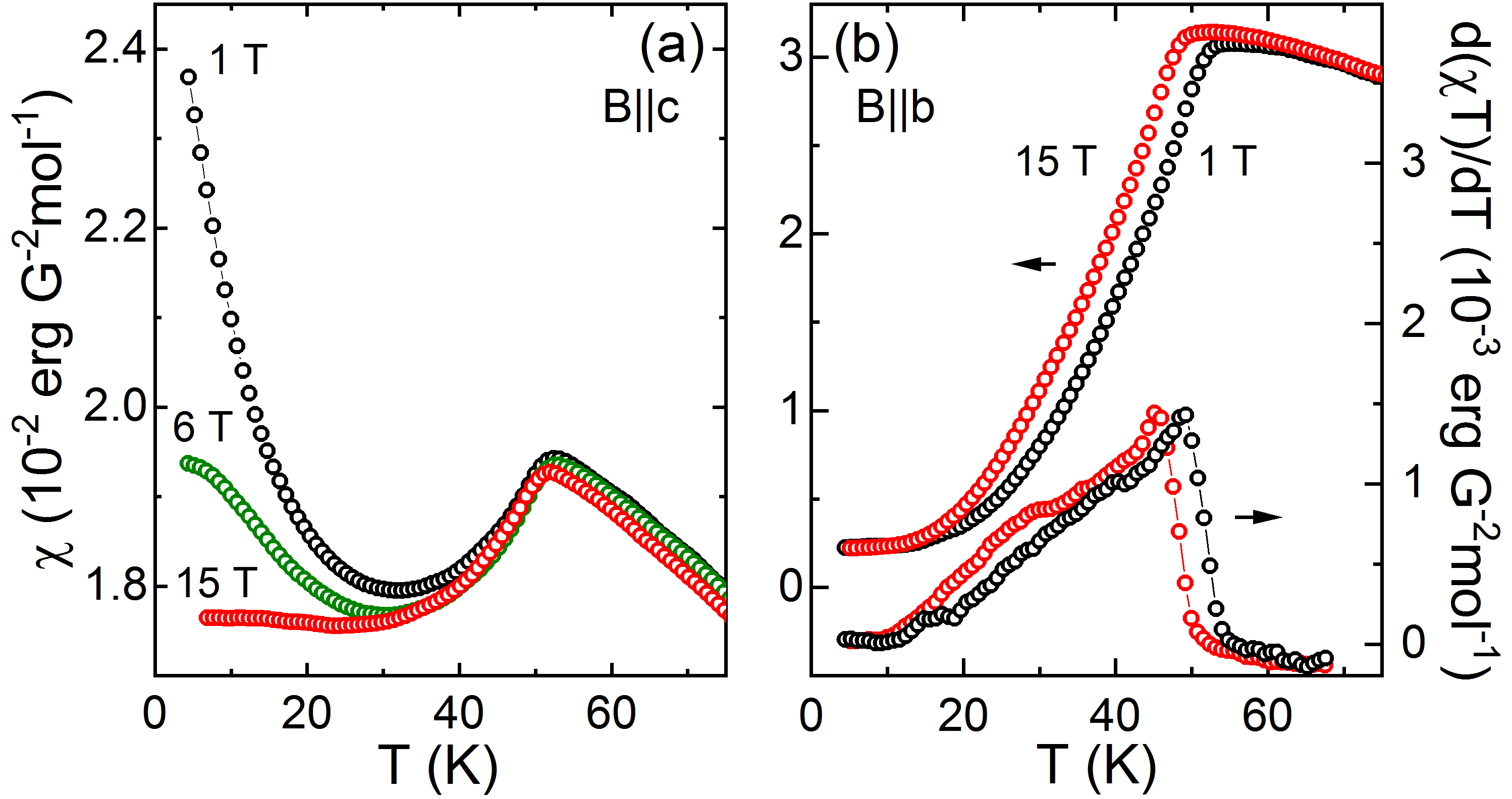}
\caption{Static magnetic susceptibility $\chi$ at $T<80$~K for (a) $B||c$-axis at 1, 6, and 15~T. (b) $\chi$ and $\partial (\chi T)/ \partial T$ for $B||b$.}\label{fig2}
\end{figure}

Application of external magnetic fields suppresses the long-range antiferromagnetically ordered phase, as visible by the effects of $B=15$~T applied along the $b$- and $c$-axis on the magnetic susceptibility shown in Fig.~\ref{fig2}. Sharp anomalies of the thermal expansion coefficients studied in various magnetic fields applied along the three crystallographic axes (Fig.~\ref{fig:LT}) enable detailed determination of the phase boundaries. These measurements are backed-up by magnetostriction data at various temperatures (see Fig.~S2 of the supplement). While for $B\|a$- and $c$-axis, neither \tn\ nor the shape of the anomalies in $\alpha$ are significantly affected by magnetic fields up to 15~T, there is a more sizeable effect for $B\|b$. Specifically, \tn\ is shifted at $B=15$~T by $\Delta T_{\rm N}\approx 3$~K (see Fig.~\ref{fig:LT}b) while the shape of the anomalies in \alb\ is only very weakly affected. This observation corresponds to the effect of $B\|b = 15$~T on Fisher's specific heat~\cite{fisher1962relation} $\partial (\chi T)/ \partial T$ presented in Fig.~\ref{fig2}b.

Notably, at $B=0$~T, the thermal expansion coefficients \alb\ and \alc\ exhibit an additional feature in the ordered phase, i.e., at $T<$~\tn . As illustrated in Fig.~\ref{alpha}a, the length changes towards lowest temperatures undergo a minimum at approximately 43~K followed by a broad step-like increase. Correspondingly, there is a minimum in $\alpha$ (\tss , open triangles in Fig.~\ref{alpha}) followed by a step-like feature (\ts , filled triangles). Note that a similar feature may be present in \ala\ but masked by the large anomaly at \tn . Qualitatively, the data in Fig.~\ref{fig:LT}b and c imply suppression of the associated phenomenon in applied magnetic fields, as indicated by reduction of the characteristic temperatures \tss\ and \ts\ as well as of the minimum- and step-size. For a quantitative estimate of the field effect, in addition to the temperature of the minimum \tss\ we extract a characteristic temperature \ts\ at the inflection point of \alb\ which shifts from \ts $(0~\mathrm{T})\approx 19$~K to about 11~K at $B\|b=15$~T.


\subsection{Magnetisation}

\begin{figure}
\includegraphics[width=1.0\columnwidth,clip] {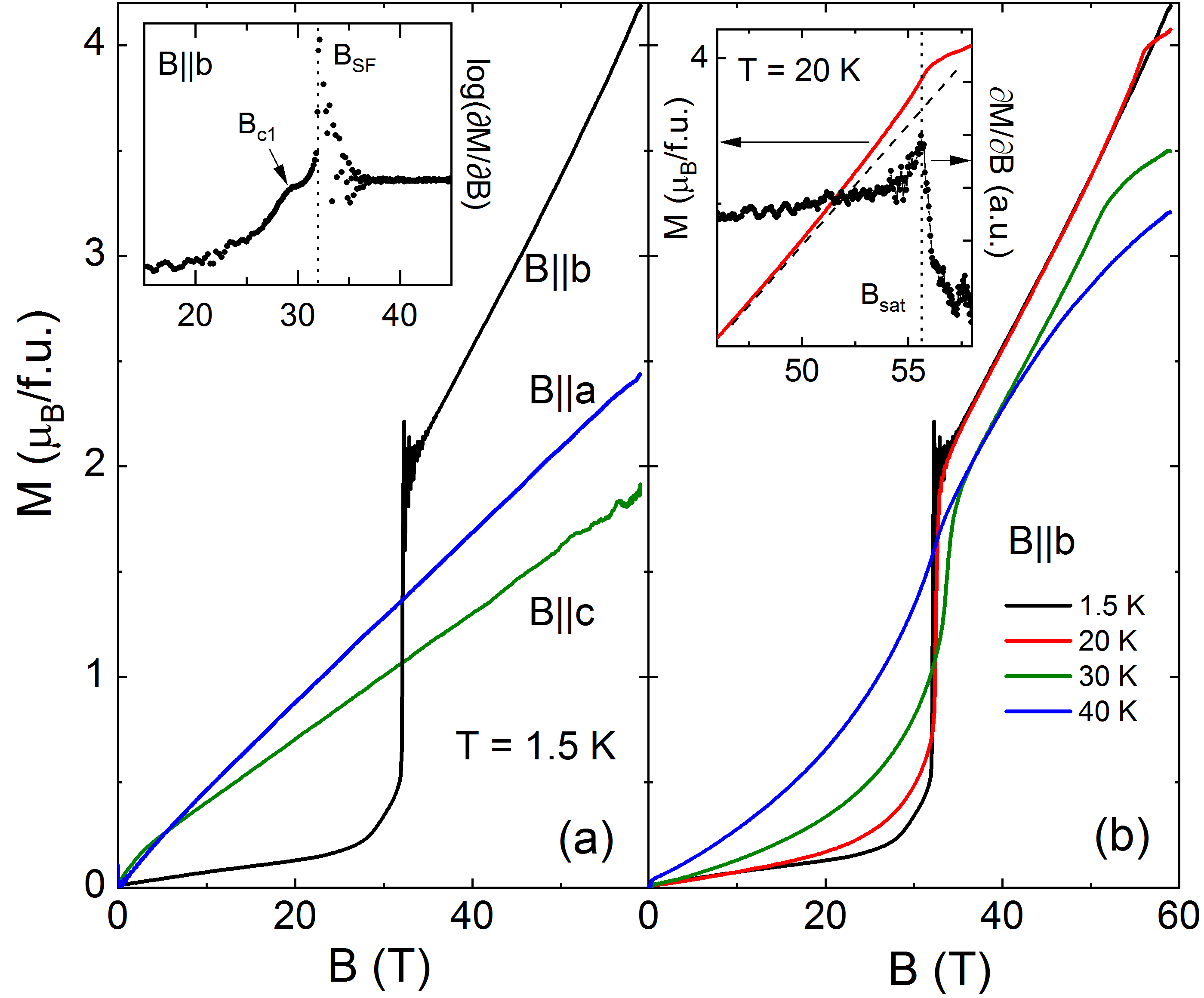}
\caption{(a) Pulsed-field magnetisation for all three crystallographic directions of \lfp , at 1.5~K, and (b) for $B||b$-axis at various temperatures $1.5\leq T\leq 40$~K. All data have been obtained upon up-sweep of $B$. Insets: (a) Magnetic susceptibility, at 1.5~K, for $B||b$ (logarithmic scale) highlighting spin reorientation as well as an anomaly at \bco . (b) Magnetisation and magnetic susceptibility in the vicinity of the saturation field \bs . The dashed line shows a linear extrapolation of the data. }\label{fig:mb}
\end{figure}

The magnetisation $M$ vs. $B$ at $T=1.5$~K along the three crystallographic axes is shown in Fig.~\ref{fig:mb}. At $T=1.5$~K, there is a jump-like increase of $M$($B||b$) suggesting spin-reorientation at $B_{\rm SF} = 32$~T in accord with the easy axis inferred from Fig.~\ref{chi}. The anomaly amounts to $\Delta M=1.39(3)$~\mb/f.u.. Note that in the two-sublattice model presented below, $\Delta M$ corresponds to a change of the angle between the spins from antiparallel to about 145$^\circ$. For $B\|b>$~\bsf , $M$ increases linearly, which is also observed for $M$($B\perp b$). At $T=1.5$~K, none of the high-field magnetisation curves show signatures of saturation up to 58~T. At small magnetic fields, there is a right-bending of the magnetisation curves which implies the presence of quasi-free spins. Notably, this behaviour significantly depends on the magnetic field direction. It is most pronounced for $B\|c$, in agreement with the Curie-like contribution to $\chi$($T$) which is largest for this field direction (cf. Fig.~\ref{chi}). Quantitatively, fitting the magnetisation curves by a Brillouin function $B_{1/2}$ plus a linear term describes the data for $B\perp c$ very well. The data indicate $M_{\rm qf \perp c}\approx 0.08$~\mb /f.u. for the response associated with quasi-free (qf) spins. Whereas, the curvature seen in $M$ vs. $B||c$ suggests much larger moments or a strongly anisotropic $g$-factor. The behaviour in $M$ vs. $B||c$ agrees to the effect on $\chi$ vs. $T$. As shown in Fig.~\ref{fig2}a, the Curie-like upturn at $T<10$~K is most pronounced for $B||c$ but completely suppressed at $B = 15$~T.

The rather linear behaviour $M$($B||b>B_{\rm SF}$) does not extrapolate to the origin of the graph. Hence, while the transition may be attributed to spin-reorientation, it is not associated with a simple spin-flop behaviour. This is corroborated by a more detailed inspection of the anomaly, as highlighted by the susceptibility $\partial M/\partial B$ in Fig.~\ref{fig:mb}. In addition to the jump at \bsf , there is a precursing broad step-like increase of $\partial M/\partial B$ towards a small plateau. The small step extends from $\sim 26$ to 29~T (the latter is labelled \bco\ in the inset of Fig.~\ref{fig:mb}a). All non-linear changes associated with \bco\ and \bsf\ sum up to $\Delta M=1.74(5)$~\mb/f.u. While there is no visible hysteresis for neither $B\perp b$ nor $B\|b>B_{\rm SF}$, a small but sizeable difference between magnetisation in the up- and down-sweeps (data not shown) confirms the discontinuous nature of the spin-reorientation process at \bsf .

Upon heating, the anomaly at \bco\ vanishes. The magnetisation jump at \bsf\ decreases and is smeared out while the critical field changes only very weakly (Fig.~\ref{fig:mb}b). At the same time, at higher temperatures the saturation field appears to be visible in the accessible field range. At $T=20$~K, we find \bs\ = 56(3)~T which is well identified by a peak in $\partial M/\partial B$ (see the inset of Fig.~\ref{fig:mb}b). In addition, there is a slight left-bending of $M$ vs. $B$ just below \bs . However, the almost constant slope of the $M$ vs. $B$ curve evidences a predominant 3D character of magnetism, which is in accord with the size of the ordered magnetic moment observed in the ordered phase at low temperatures.~\cite{toft2015anomalous,nishimoto2011}


\subsection{Magnetic phase diagram}

\begin{figure}
\includegraphics[width=1.0\columnwidth,clip] {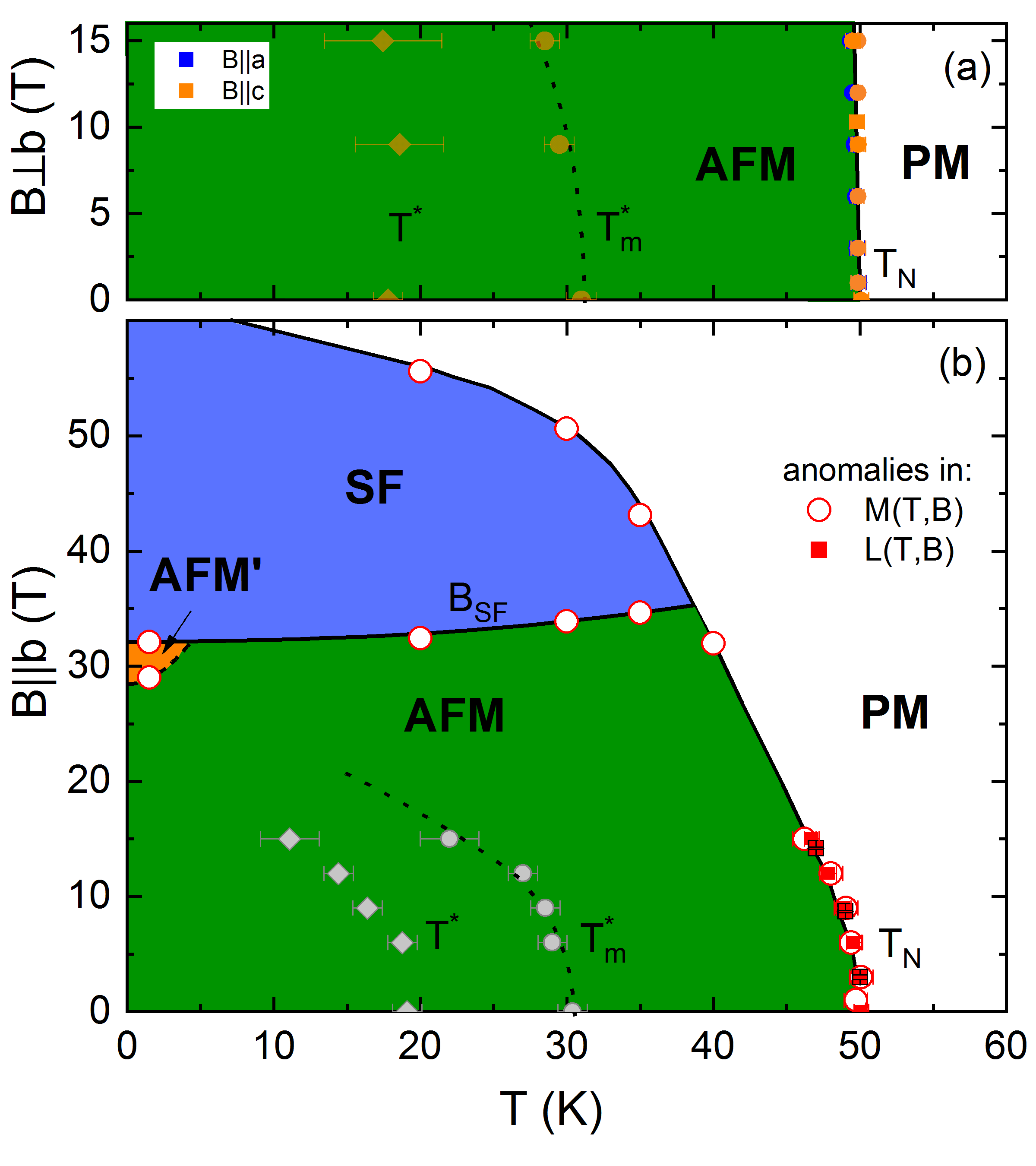}
\caption{Magnetic phase diagram of \lfp\ for (a) $B\|a$ (blue markers), $B\|c$ (orange) and (b) $B\|b$ (red) as constructed from thermal expansion/magnetostriction (squares) and magnetisation (circles) measurements. The lines are guides to the eye. AFM, AFM', SF, and PM label antiferromagnetically ordered, spin-reoriented, and paramagnetic phases, respectively. \tn and \bsf\ label the associated anomaly temperatures and fields. \ts\ (\tss) labels the step-like (minimum) feature in $\alpha$ (grey markers; see text).}\label{fig:PD}
\end{figure}

The magnetic phase diagram shown in Fig.~\ref{fig:PD} summarizes the evolution of the anomalies observed in the thermal expansion, magnetostriction, and magnetisation upon application of external magnetic fields. For $B\perp b$, the data display the anomaly at \tn ($B$) which is only weakly field-dependent. In particular, the phase boundaries \tn $(B)$ for fields parallel to the $a$- and $c$-axis overlap, and magnetisation, specific heat and thermal expansion measurements show $\partial T_{\rm N}/\operatorname{\partial B}=0.023(8)$~K/T.

For $B\|b$-axis, a much more pronounced field effect is observed. For small $B$, i.e., in the vicinity of \tn , $\partial T_{\rm N}/\operatorname{\partial B}=0.083(4)$~K/T is revealed. Extrapolating the phase boundary \tn ($B$) to low temperatures suggests \bs ($B||b$) = 64(2)~T. In addition, at $B_{\rm SF}$($T$=1.5~K) $ \approx 32.0(1)$~T, a spin-reoriented phase (SF) evolves. \bsf\ is almost temperature-independent. A rough estimate by means of a Clausius-Clapeyron relation $\Delta S=-\Delta M\cdot\partial B/\partial T \approx 0.15$~\jmk\ implies only insignificant entropy changes associated with this transition.~\cite{stockert2012pr} At $T=1.5$~K, there is a precursing anomaly in $M$($B$) indicating the presence of a competing antiferromagnetic phase AFM' evolving at \bco\ = 29~T. Finally, the phase diagram in Fig.~\ref{fig:PD} presents characteristic temperatures/fields associated with the above-discussed feature in the thermal expansion coefficient which signals a structure-dielectric coupling. Fig.~\ref{fig:PD} displays the characteristic temperatures \ts\ and \tss\ of the step-like behaviour and the minimum in $\alpha$, respectively.

\section{Discussion}

\begin{figure}
	\includegraphics[width=1.0\columnwidth,clip] {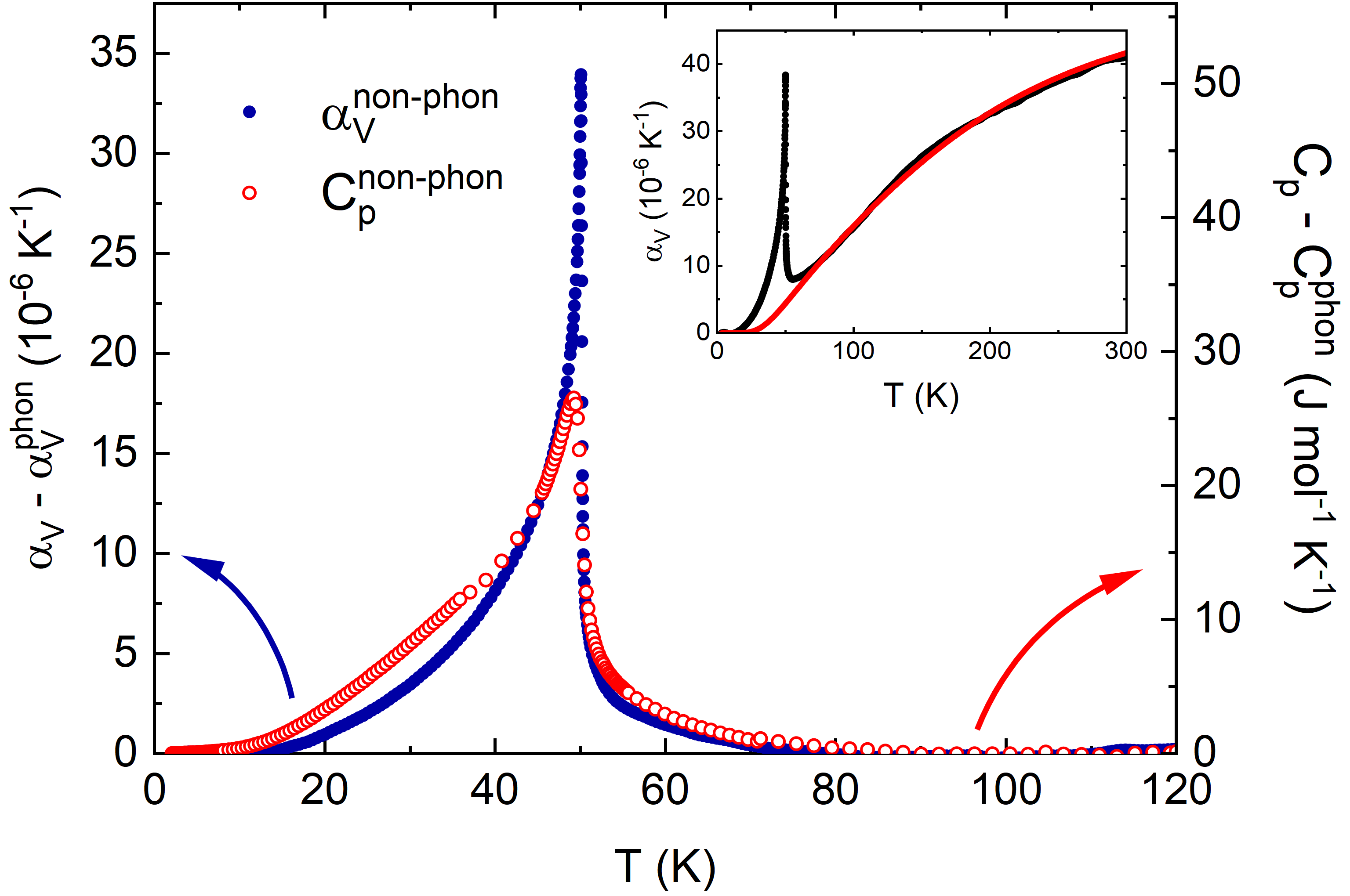}
	\caption{Gr\"{u}neisen scaling of the non-phononic contributions to the heat capacity (uncorrected \cp\ data have been taken from Ref.~\onlinecite{loos2015heat}) and volume thermal expansion coefficient. The inset shows the volume thermal expansion coefficient \alv\ (black markers) with a combined Debye-Einstein-fit at high temperatures (red line).}\label{fig:grueneisen}
\end{figure}

Comparing the non-phononic contributions to the specific heat and to the thermal expansion coefficient enables further conclusions on the nature of the associated ordering phenomena. In order to clarify the presence of one or more relevant energy scales, the volume thermal expansion coefficient \alv\ (Fig.~\ref{fig:grueneisen} inset), as derived by adding the uniaxial coefficients $\alpha_i$, is to be compared with the respective entropy changes, as measured by the specific heat. For this comparison we use specific heat data by Loos \etal ~\cite{loos2015heat} obtained on polycrystalline \lfp . To be specific, for a comprehensive Gr\"{u}neisen analysis, the lattice contributions of both \alv\ and \cp\ have to be separated. Extending the analysis of lattice contributions to \cp\ in Ref.~\onlinecite{loos2015heat}, \alv\ and \cp\ are simultaneously fitted at temperatures well above the magnetic anomalies by means of a combined model consisting of a sum of Debye and Einstein terms:

\begin{align}
c_{\rm p}^{ph} &= &&n_D \cdot D(\Theta_D/T) + &&n_E \cdot E(\Theta_E/T) \\
\alpha_{\rm V}^{ph} &= ~\gamma_D &&n_D \cdot D(\Theta_D/T) + ~\gamma_E &&n_E \cdot E(\Theta_E/T)
\end{align}\label{eq:debye}

Debye- $D$ and Einstein-function $E$ depend on their characteristic temperatures $\Theta_D$ and $\Theta_E$, respectively. $n_{\rm D}$ and $n_{\rm E}$ denote the number of modes associated with each contribution such that their sum yields the number of atoms in the unit cell. The use of two Gr\"{u}neisen parameters $\gamma_D$ and $\gamma_E$ accounts for different Gr\"{u}neisen scalings for the individual summands and is necessary, as the contributions cannot be treated separately in the investigated temperature range. The fit for \alv\ is depicted together with the measured data in the inset of Fig.~\ref{fig:grueneisen}. The procedure yields a good description of the high temperature behaviour with $n_{\rm D}=3.77$~mol$^{-1}$, $n_{\rm E}=2.29$~mol$^{-1}$, $\Theta_D =833$~K, $\Theta_E =229$~K. These values are consistent with the previous analysis of the specific heat data~\cite{loos2015heat}. The Gr\"{u}neisen parameters amount to $\gamma_D = 3.81\cdot 10^{-7}$~mol/J and $\gamma_E = 3.06\cdot 10^{-7}$~mol/J. Due to the fact that the present analysis employs a concomitant fit of both the length and entropy changes, it may be valuable to report the resulting entropy changes $S_{\rm mag}=12.4$~\jmk\ obtained by integrating the data by Loos \etal\ corrected by the hereby obtained phononic background. This value is larger than in Ref.~\onlinecite{loos2015heat} and closer to the theoretically expected value of 13.38~\jmk.

Comparison of the non-phononic parts of \cp\ and \alv , i.e., the respective differences of the measured data to the phononic fits, enables investigating the Gr\"{u}neisen ratio of the associated length and entropy changes. Accordingly, the (magnetic) thermal expansion coefficient and the specific heat are shown in the main of Fig.~\ref{fig:grueneisen}. Firstly, the data imply that the above mentioned procedure yields reliable results as there is a large temperature regime where \cpm\ and \alm\ are proportional to each other. The experimental data and their analysis hence clearly show that the entropy and length changes in this temperature regime are driven by one degree of freedom. It is tempting to attribute this mainly to the spin degrees of freedom, i.e., the entropy changes are of magnetic nature which is supported by the fact that the extracted non-phononic entropy changes $\Delta S$ nearly agree to the expected spin entropy. To be specific, while there is no magnetic contribution to \cp\ and \alv\ above $\sim 90$~K, the two quantities match well down to about 40~K, including the behaviour at \tn . Scaling yields a Gr\"{u}neisen parameter of $\gamma_{\rm mag}=6.7(5)\cdot 10^{-7}$~mol/J. This value is associated with the pressure dependence of \tn\ being $dT_{\rm N}/dp = \gamma_{mag} T_{\rm N} V_{m} = 1.5(1)$~K/GPa which is deduced using the molar volume $V_{m} = 43.6$~cm$^{3}$.~\cite{neef2017high}

Notably, however, the scaled data show a significant deviation from each other well below \tn\ at temperatures between roughly 10 and 45~K, i.e., Gr\"{u}neisen scaling by means of $\gamma_{\rm mag}$ is not valid in this temperature regime. In general, failure of Gr\"{u}neisen scaling implies the presence of additional relevant degrees of freedom. Here, it provides thermodynamic evidence that several dominant degrees of freedom are concomitantly relevant in the ordered phase.
Phenomenologically, failure of Gr\"{u}neisen scaling can be associated with the low temperature upturn in length changes of the $b$- and $c$-axis, respectively, upon reducing the temperature. It also agrees to the temperature regime below the peak of the magneto-electric coupling coefficient \axy .~\cite{toft2015anomalous} One must conclude that the low-temperature feature observed in the thermal expansion must be ascribed to one or more additional degrees of freedom not corresponding to only magnetic entropy which drives antiferromagnetic order at \tn .

One may speculate about linking the observed failure of Gr\"{u}neisen scaling to spectroscopic properties of \lfp . Yiu \etal ~\cite{yiu2017hybrid} have detected rather dispersionless hybrid excitations which are discussed in terms of electron states arising from the crystal-field splitting and spin-orbit coupling. Employing the parameters from Ref.~\onlinecite{yiu2017hybrid} for a Schottky-like model would imply broad humps in the specific heat and the thermal expansion coefficient which are centered around 32~K. As pressure dependence of the underlying energy scale of this phenomenon is supposed to be different from that driving long-range spin order, one must expect a different Gr\"{u}neisen parameter for this expected hump. Failure of Gr\"{u}neisen scaling in \lfp\ may hence be straightforwardly associated with the reported hybrid excitations.


The measured spin-flop field at 1.5~K, and the extrapolated saturation field \bs ($B||b$) = 64(2)~T allow to determine the effective antiferromagnetic exchange interaction $J_{\rm AF}$ and the anisotropy difference $D_{\rm b}$ between the easy axis ($b$-axis) and the intermediate axis ($a$-axis) in a two-sublattice mean-field model. It is described by

\begin{align}
	\mathcal{H} = &J_{\rm AF} S_i \cdot S_j + D_{\rm b}(S_i^b)^2 + g \mu_{\rm B} B \cdot (S_i + S_j) \label{eq:ham}
\end{align}

with the magnetic field $B||b$, $g_b = 2.31$, $S^b$ the spin component in $b$-direction and $\mu_{\rm B}$ the Bohr magneton. $J_{\rm AF}$ is the effective exchange interaction between the sublattices $i$ and $j$.  Inter-sublattice exchange interactions are not considered for this analysis. The model yields $J_{\rm AF} = 2.68(5)$~meV and $D_{\rm b} = -0.53(4)$~meV. The minus sign of $D$ signals that at $B = 0$~T, spins align along the $b$-axis. Extending the Hamiltonian by an additional anisotropy in $c$-direction, i.e., $D_{\rm c}$ similar to $D_{\rm b}$, enables to account for the different susceptibilities $\partial M/\partial B$ measured along the $a$- and $c$-axis (see Fig. \ref{fig:mb}a). Quantitatively, we obtain the plane-type anisotropy $D_{\rm c} = 0.44(8)$~meV.

Although the Hamiltonian (Eq.~\ref{eq:ham}) provides only a basic model for evaluating magnetism in \lfp\ which neither covers the two-step nature of spin-reorientation (i.e., the presence of the intermediate phase AFM') nor takes into account that $M$($B>$\bsf ) does not resemble a simple spin-flop scenario, the obtained anisotropy of $D_{\rm b} = -0.53(4)$~meV is in good agreement to values obtained by inelastic neutron scattering (INS) where -0.62(2)~meV\cite{li2006spin} and -0.86(1)~meV \cite{yiu2017hybrid} have been reported.~\footnote{In Refs.~\onlinecite{yiu2017hybrid,li2006spin}, the anisotropy tensor was chosen such that $D_{\rm b} = 0$, whereas the present work uses a notation with $D_{\rm a} = 0$. Therefore, cited values have been converted into the notation of the present work.} Moreover, the effective exchange interaction  $J_{\rm AF} = 2.68(5)$~meV deduced from the macroscopic data at hand is in a good agreement with $J_{\rm AF} = 4(J_{\rm bc}+J_{\rm ab}) = 3.64(2)$~meV~\cite{li2006spin} and $2.20(6)$~meV~\cite{yiu2017hybrid} from INS. On the other hand, $D_{\rm c} = 0.44(8)$~meV does not agree with the INS results 0.94(4)~meV~\cite{li2006spin} and 1.37(2)~meV~\cite{yiu2017hybrid}.~\footnote{Note that the actual numbers depend on the $g$-factors used in the analyses. In the present work, $g$-values from high-temperature Curie-Weiss fitting is used while the INS models employ $g = 2$.}

The low-temperature upturn of static susceptibility as well as the right-bending of the magnetisation curves below $B = 10$~T indicate the presence of anisotropic quasi-free moments. It has been shown~\cite{neef2017high} for single-crystals LiMn$_{1-x}$Fe$_x$PO$_4$ that the presence of such anisotropic moments evolves with increasing iron content $x$. One may speculate that the quasi-free moments originate from Fe$^{2+}$ anti-site disorder~\cite{gardiner2009anti,neef2017high}, whereby some Fe$^{2+}$-ions reside on Li-positions due to the similar radii of the two respective ions. Anti-site disorder in the investigated crystal has been estimated to about $2.3(2)\%$~\cite{neef2017high}.

\section{Summary}

The reported experimental studies of pulsed- and static-field magnetisation, thermal expansion, and magnetostriction of single-crystalline \lfp\ enable constructing the magnetic phase diagram. In addition, high-resolution dilatometry is used for quantitative analysis of the pronounced magneto-elastic coupling in \lfp . The macroscopic data imply antiferromagnetic correlations well above \tn, up to about 250~K. This is corroborated by observation of magnetic contributions to the thermal expansion which obey Gr\"{u}neisen scaling far above \tn . Notably, recently reported temperature dependence of the magneto-electric coupling coefficient \axy\ [\onlinecite{toft2015anomalous}] is linked to the failure of Gr\"{u}neisen scaling. Our data hence provide direct thermodynamic experimental evidence for the essential role of structure changes for magneto-electric coupling in \lfp . Upon application of magnetic fields, associated features are suppressed. In addition, for $B||b$-axis and $T=1.5$~K, a pronounced jump in the magnetisation implies spin-reorientation at $B_{\rm SF} = 32$~T as well as a precursing competing phase at 29~T. In a two-sublattice mean-field model, the saturation field $B_{\rm sat,b} = 64(2)$~T and the spin-flop field $B_\mathrm{SF} = 32.0(1)$ enable the determination of the effective antiferromagnetic exchange interaction $J_{\rm af} = 2.68(5)$~meV as well as the anisotropies $D_{\rm b} = -0.53(4)$~meV and $D_{\rm c} = 0.44(8)$~meV.

\begin{acknowledgements}
The project is supported by Deutsche Forschungsgemeinschaft (DFG) through KL 1824/13-1. We acknowledge the support of the HLD at HZDR, member of the European Magnetic Field Laboratory (EMFL). JW acknowledges support from the HGSFP and IMPRS-QD. SS acknowledges support by DFG through KL 1824/6.
\end{acknowledgements}

\end{document}